# Conductance, continuity, and ferromagnetic percolation thresholds in thin films

G. Kopnov and A. Gerber

School of Physics and Astronomy
Tel Aviv University
Ramat Aviv 69978 Tel Aviv, Israel

Abstract

Classical percolation models predict the metal-insulator transition and the onset of the long-range ferromagnetic order at the same topological continuity threshold. We tested this prediction in thin films of ferromagnetic CoPd and found a dramatic difference between the conductance and magnetic thresholds. While the long-range ferromagnetic phase develops at or very close to the continuity threshold, the transition from the metal-like to insulator-like conductance develops in films several times thinner. We argue that atomically narrow low resistance gaps intersecting the fractal network of metallic clusters provide a consistent explanation of the effect. We identify the conduction threshold as the point in discontinuous films at which the resistance of intergranular junctions exceeds the quantum resistance mark.

Introduction

The correlation between the topology and electrical conductivity of thin films and metal-insulator granular systems attracts long-standing interest. Classical percolation models [1, 2] assume zero conductance for any discontinuity in metallic networks and predict the conductivity to follow the scaling law:

$$\sigma \propto (x - x_c)^\gamma \qquad (1)$$

where $x$ is the fractional volume of the conducting phase in three-dimensional (3D) systems or its planar coverage in two-dimensional (2D) films, $\gamma$ is the universal transport exponent ($\gamma \approx 1.3$ and 2 in 2D and 3D systems respectively), and $x_c$ is the continuity threshold. Such scaling has been observed in many systems in the vicinity of the metal-insulator transition [3, 4]. However, identification of the continuity threshold with the onset of conductivity can be not accurate due to quantum and thermally assisted tunneling across the discontinuity gaps. It has been found that the tunneling conductance in several granular compounds [5, 6] follows a similar scaling behavior (Eq. 1) well beyond the metallic particles' coalescence concentration. The exponents of this tunneling-type scaling can differ significantly from the universal metallic-type percolation. $\gamma$ as large as 6.4 was found in carbon black – polymer composites [7]. The questions have been asked regarding the onset of tunneling conductivity, its proper description, and physical meaning of the experimentally observed scaling and critical thresholds $x_c$ [8 - 10].

The property regularly used to distinguish between continuous and discontinuous metallic structures is a reversal of the resistivity-temperature coefficient (RTC) from positive in continuous to negative in discontinuous networks where the conductance is driven by thermally activated tunneling [11]. Such classification assumes that resistance of tunnel junctions immediately below the continuity threshold is dominant in the conducting circuit and is reflected in the measured data. This may depend on granular topology. Thin films fabricated by conventional deposition techniques on insulating substrates usually have maze patterns with labyrinths of metallic clusters and intergranular voids. Resistance of intergranular gaps can vary greatly depending on temperature and junction geometry (cross-section vs width). Resistance of atomically narrow discontinuities at non-zero temperatures can be lower than that of metallic clusters they separate, such that their insulator-like temperature dependence is hidden behind the dominant metallic one [12]. The negative RTC at e.g. room temperature develops when intergranular gaps are sufficiently wide, which occurs at film coverage lower than the geometrical connectivity threshold. How much lower is an open question?

We try to resolve this puzzle by using another phenomenon taking place at the topological continuity threshold: the onset of long-range ferromagnetic order. Below the continuity threshold, granular magnetic films are composed of finite-size paramagnetic or superparamagnetic clusters separated by voids in thin films or embedded within an insulating host, such as $SiO_2$ in 3D mixtures. The crystalline size, topology, and film growth mechanism depend on the material, substrates, and fabrication conditions [13 – 19]. Two scenarios of topology-related ferromagnetic ordering in granular systems were discussed in the literature. The first mechanism is direct exchange interaction. In this case, the transition from the para/superparamagnetic to the ferromagnetic phase is ascribed to the formation of an infinitely large physically continuous magnetic cluster spreading over an entire film surface. Consequently, the ferromagnetic percolation threshold is identical to the continuity one [20, 21]. The second possibility is the development of ferromagnetic ordering among close yet detached magnetic clusters by sufficiently strong intergranular interactions such as the dipole-dipole one [22, 23, 24]. Dipolar interactions are highly anisotropic and depend on the particle arrangement. One can expect the dipolar ferromagnetic ordering in 1-D chains of nanoparticles or 2-D arrays of nanoparticles with in-plane magnetic anisotropy [25]. It is less likely in films with uniaxial perpendicular to plane magnetic anisotropy, such as CoPd used in this study, in which case the stray field is expected to promote the anti-ferromagnetic rather than the ferromagnetic order. In any case, the ferromagnetic ordering takes place at or below the continuity percolation threshold and can mark its lowest boundary. Here, we adapted the onset of the long-range ferromagnetic phase in thin CoPd films as the continuity threshold. As will be shown, the difference between the continuity/ferromagnetic threshold and the conductance one is dramatic. The metal-like conductance is preserved in discontinuous films up to four times thinner than the continuity threshold.

Experimental

The films studied here are of $Co_{20}Pd_{80}$ alloy. In bulk, the Curie temperature of this alloy is about 500 K, and the magnetic moment per Co atom is 3.5 $\mu_B$ [26]. Thick films exhibit out-of-plane magnetic anisotropy due to strong negative magnetostriction coefficient [27]. The material was studied for potential use in magnetic random-access memory [28, 29] and the Hall effect spintronic detection of hydrogen [30]. It was also found that magnetic properties are highly sensitive to the film thickness in the vicinity of the metal-insulator transition [31]. As will be shown in the following, thin films of this alloy are particularly suitable for the study of the topology-driven ferromagnetic ordering at room temperature. Polycrystalline $Co_{20}Pd_{80}$ (atomic concentration) films with thickness from 1.5 nm to 100 nm and lateral dimensions 5 × 5 mm were fabricated by co-sputtering from separate Co and Pd targets at $5\times10^{-3}$ mbar Ar pressure on polished semi-insulating GaAs substrates (resistivity higher than $10^7 \Omega$cm) held at room temperature. The composition was controlled by the relative rf-power of the respective sputtering sources and tested by the energy dispersive x-ray spectroscopy analysis (EDAX). No post-deposition annealing was made. The film thickness used in the text is the average value defined as the total mass deposited per unit area divided by the bulk density. The thickness was calibrated using the Hall bar samples scanned with the atomic force microscope (AFM) Park Systems NX10. Transmission electron microscopy (TEM) images were taken with JEOL JEM-2010F UHR device. The extraordinary Hall effect (EHE) was used for magnetic characterization. EHE is proportional to the out-of-plane magnetization and is appropriate for the study of ultrathin magnetic films [32]. Resistance, the ordinary (OHE), and extraordinary (EHE) Hall effects were measured as a function of temperature and magnetic field using the Van der Pauw protocol.

Results and discussion

Fig.1 illustrates the polycrystalline structure and morphology of the deposited films. Individual fcc crystallites with pronounced (111) out-of-plane growth texture and random in-plane lattice orientations are seen in the high-resolution TEM micrograph (Fig.1c). The lateral crystallite dimensions are 3 – 5 nm. Figs. 1a and 1b illustrate the morphology of the 1.5 nm and 7 nm thick films. CoPd (dark in the figure) forms a typical percolation maze pattern. The 1.5 nm thick film is obviously below the continuity threshold. Metallic clusters are finite-size surrounded by gaps 1.5 - 2 nm wide. As many other polycrystalline films deposited on insulating substrates, the growth follows the Volmer-Weber mode [33, 34] in which clusters expand both laterally and vertically when more material is added, and the coverage gets denser with increasing thickness. By binarizing Figs. 1a and 1b we found the planar coverage of the 1.5 nm and 7 nm thick films to be approximately 75% and 88%, respectively. The addition of 5.5 nm material increased the lateral coverage by 13% only, while the average height of CoPd clusters in the 7 nm film grew about 4.1 times higher than in the 1.5 nm one. However, visual inspection of the 7 nm thick sample cannot determine whether the metal is continuous on a macroscopic scale or narrow void channels within the layer form an infinite network dividing the metal into finite disconnected clusters. One should keep in mind that the topology of films deposited on amorphous carbon grids for TEM microscopy can differentiate from the samples grown on crystalline GaAs substrates used for transport measurements. However, earlier studies of CoPd films grown on amorphous glass and crystalline GaAs did not reveal significant differences between the substrates [30].

The topological percolation threshold is defined as the point at which the size of metallic clusters diverges toward infinity. The saturated magnetic moments of finite-size ferromagnetic clusters are proportional to their volume. At temperatures below the Curie and above the blocking, the system is superparamagnetic, and its field-dependent magnetization can be described by the Langevin function. This allows an accurate determination of the effective magnetic moment and, respectively, the cluster size. Fig.2 presents the normalized EHE resistance of a series of $Co_{20}Pd_{80}$ samples with different thicknesses as a function of the normal-to-plane magnetic field at room temperature. Magnetization of all films thinner than 7 nm is hysteresis–free. Minor hysteresis develops in 8 nm thick film on the S-shape background. Films thicker than 10 nm exhibited sharp magnetization reversal with square hysteresis and close to full magnetic remanence. The hysteresis-free magnetization of thin films was fitted by the Langevin function:

$$L(J) = cothJ - 1/J \qquad (2)$$

where $J = \mu H/k_B T$, and $\mu$ is the moment of the effective cluster. Solid lines in Fig.2 are fitting of the data by Eq. 2. The effective magnetic moment $\mu$ extracted from the fitting is plotted in Fig. 3 as a function of film thickness. The smallest magnetic moment $\mu = 1600\mu_B$ was calculated for the thinnest 3.5 nm thick film. This value corresponds to a cubic grain of about 3.2 nm in size [35], which is in good agreement with the individual crystallite dimensions found by the high-resolution microscopy (Fig. 1a). The moment grows sharply above 6 nm thickness. The thickness dependence can be fitted by the function:

$$\mu = \mu_0(t_{cFM} - t)^\alpha \qquad (3)$$

with the critical magnetic thickness $t_{cFM} = 7.7 \pm 0.1$ nm and power index $\alpha$ = -2.2 $\pm$ 0.2. This power index is close to the critical exponent 2.4 predicted for the scaling of cluster size in the 2D percolation phase transition [1, 2]. The divergence of the magnetic moment and development of robust hysteresis beyond the critical thickness can be interpreted as the 2-dimensional superparamagnetic to ferromagnetic phase transition. Since the moment is proportional to the effective cluster size, the latter diverges at $t_c = t_{cFM}$. Thus, the topological continuity threshold is found at $t_c = 7.7 \pm 0.1$ nm.

Similar scaling divergence of magnetic susceptibility was observed in other thin film materials and interpreted as the topological onset of the long-range ferromagnetic phase. Epitaxial films grown on single-crystalline metallic substrates such as Co/Cu [13, 14], CoFe/Cu [15], and Fe/W [16] were considered two-dimensional and paramagnetic at low film coverage at room temperature. Films deposited on insulating substrates, such as Fe on GaAs or InAs [17, 18] and Co on $Al_2O_3$ [19] formed superparamagnetic polycrystalline island structures at early stages of growth. In all cases, the transition from the para/superparamagnetic to the ferromagnetic phase was ascribed to the formation of an infinitely large physically continuous magnetic cluster spreading over an entire film surface.

The thickness dependence of the planar resistance $R_\square$ is plotted in the same Fig. 3 (left vertical axis). The resistance increases gradually with decreasing thickness and diverges in the thin film limit as:

$$R_\square = R^*(t - t_{cR})^{-\gamma} \tag{4}$$

with $\gamma = 1.2 \pm 0.15$ and the resistivity critical threshold $t_{cR} = 2 \pm 0.2$ nm. The log-log presentation of $R_\square$ as a function of $t - t_{cR}$ is shown in the inset. Such scaling with the universal power index $\gamma \approx 1.3$ is consistent with the 2D conductance percolation theory (Eq.1). As mentioned above, classical percolation models predict the metal-insulator transition and the onset of the long-range ferromagnetic order at the same topological continuity threshold. Remarkably, resistance follows the classical conductance scaling (Eq.1) with the predicted universal power index but the critical thickness $t_{cR}$ is about four times smaller than the continuity threshold $t_c = t_{cFM}$

Identification of the continuity threshold at $t_{cFM}$ is supported by more experimental evidence. Magnetoresistance can serve as an independent test of the continuity or discontinuity of ferromagnetic networks. Two types of magnetoresistance behavior are expected in ferromagnetic materials in the vicinity of the topological continuity threshold. Anisotropic magnetoresistance (AMR) [36] is expected to be the main mechanism in continuous bulk-like films. The spin-dependent tunneling magnetoresistance (TMR) across the insulator gaps is dominant when films are discontinuous [37 - 39]. The qualitative difference between the two mechanisms is revealed when the magnetic field is applied in-plane parallel/antiparallel to the current direction. The AMR magnetoresistance is positive at small fields up to the magnetic saturation, while the TMR is negative in the same field range. The transition between the AMR and TMR effects has been observed in granular ferromagnet–insulator mixtures Ni-SiO$_2$, Co-SiO$_2$ [39, 40] and in thin Ni films in the vicinity of the metal–insulator transition [41]. Fig.4 presents the normalized magnetoresistance $\frac{\Delta R}{R(0)} = \frac{R(B)-R(0)}{R(0)}$ of several CoPd films with thicknesses between 4 and 10 nm measured at room temperature with a field applied parallel to the current flow. The negative slope at high fields is the "paramagnetic magnetoresistance" due to the spin-magnon scattering [42]. Sharp low field increase of resistance in 8 nm and 10 nm thick samples is a characteristic AMR signature, indicating the ferromagnetic continuum. The positive AMR is suppressed in the 7 nm thick film and is replaced by the dominant TMR in films thinner than 6 nm. The latter indicates the division of the system into multiple fragmented clusters. The change in the dominant magnetoresistance mechanism occurs across the same threshold found by the divergence of the superparamagnetic moment.

Fig. 5 presents the resistance temperature dependence of a thick 20 nm film and several films in the thickness range $t_{cR} \leq t < t_{cFM}$. The resistance of films thicker than 8 nm is metallic in the entire temperature range with a positive resistivity temperature coefficient $\alpha = dR/dT \geq 0$ saturating to zero at low temperatures. Films thinner than 2 nm are insulator-like ($\alpha < 0$ in the entire temperature range). Their resistance temperature dependence varies from logarithmic to exponential with decreasing thickness. Resistance of films with intermediate thickness (2 nm < t < 6 nm) is non-monotonic with the minimum resistance at a certain temperature $T_{min}$ followed by the logarithmic temperature dependence $\Delta R \propto \ln T$ below $T_{min}$. $T_{min}$ as a function of thickness is plotted in the inset. The overall resistance change between the room and 4.2 K is minor: $\left|\frac{R(4.2)-R(300)}{R(300)}\right| \leq 10\%$. The resistivity minimum followed by a logarithmic increase at lower temperatures has been observed in numerous 2D and 3D granular materials in the vicinity of the metal-insulator transition [12]. For years the phenomenon was attributed to the onset of weak localization or electron-electron interactions [43, 44] despite several qualitative and quantitative contradictions: 1) The logarithmic variation of conductivity was predicted by these models solely for two-dimensional systems but was found in many three-dimensional materials as well; 2) The temperature of the resistivity minimum can be above 100 K which is one – two orders of magnitude higher than the possible limit of quantum corrections; 3) The minimum is preserved under high magnetic fields [12], which contradicts the weak localization mechanism. A granular interpretation of the effect was suggested in Ref. 12. Following the model by Efetov and Tschersich [45, 46] the temperature variation of resistivity in granular materials depends on the resistance of tunnel junctions. For junctions with resistance higher than the quantum one ( $R_Q = h/2e^2 = 12.9\ k\Omega$), the temperature dependence is exponential, and the system is defined as strongly insulating. For junctions with resistance lower than the quantum, the temperature dependence is logarithmic both in two-dimensional and three-dimensional cases, and the system is defined as weakly insulating. The tunneling resistance of intergranular gaps with negative TRC in this regime is of the same order as metallic resistance of fractal branches with positive TRC, and one can expect the crossing point between the two mechanisms. Resistivity minimum at temperature $T_{min}$ was identified [12] as the transition between the intragranular and intergranular dominated regimes. By accepting this interpretation, granular films demonstrating the metal-like positive TRC at room temperature followed by a resistivity minimum at lower temperatures are below the geometrical percolation threshold. This is consistent with the conclusions of the magnetic characterization.

The huge difference between the continuity threshold and the onset of the insulator-like behavior can be explained by the existence and phenomenological importance of the transition range between the topologically continuous and discontinuous phases in which narrow low resistance gaps intersect the fractal networks of metallic clusters. Such discontinuities among crystalline metallic clusters are seen clearly in the high-resolution micrographs (Fig. 1) although their extension over macroscopic scales can't be estimated by visual inspection. Resistance of intergranular junctions can be evaluated by the planar resistance (planar resistance of a two-dimensional square array of resistors is equal to the resistor itself). $R_\square$ of the 8 nm thick film at the continuity percolation threshold is just 150 Ω, two orders of magnitude lower than the quantum resistance $R_Q$. The planar resistance of the 2.2 nm thick sample, the thinnest among the presented in Fig.3, is about 5 $k\Omega$ which corresponds to the three-dimensional resistivity $\rho = R_\square t \approx 1\ m\Omega$cm. That means that all samples analyzed in this work belong to the "metallic" range as determined by the Mooij criterion [47]. Films with resistance higher than $R_Q$ exhibit a truly insulator-like exponential resistance growth. Thus, the resistance percolation threshold $x_{cR}$ (Fig.3) indicates not the continuity threshold but the film coverage below which the resistance of junctions exceeds $R_Q$ and becomes significantly higher than the resistance of metallic fractals they separate.

The presence of narrow low resistance gaps intersecting the percolating network of metallic clusters can explain the extension of the conductance scaling well below the continuity threshold, and an unexpectedly wide thickness range over which the conductance scaling is observed. The power law of conductance scaling (Eq.1) was predicted by the percolation models for a limited narrow range of concentrations $\frac{x-x_c}{x} \ll 1$ only [48]. As seen in Fig.3 for CoPd films and in many other cases [3, 6, 49] the range of the scaling behavior is extended (2 to 10 nm) and not limited by theoretical predictions. This can be probably attributed to the vertical growth of expanding clusters when more material is added. As mentioned earlier, the average thickness of the 7 nm film (Fig.1b) is about four times larger than that of the 1.5 nm one. The films are quasi-two dimensional with clusters expanding both laterally and vertically, while an infinite network of narrow intergranular gaps is preserved up to $t_c$.

Previously, the role of low-resistance intergranular gaps was fully appreciated in granular superconductors. The global superconducting phase can be established in films below the continuity percolation threshold by Josephson tunneling mechanism across junctions with the normal state resistance lower than the quantum one (more precisely $R_Q/2$ due to the Cooper pair tunneling) [50 – 52]. The charge transfer across junctions with higher resistance is by single electron tunneling only. Such systems are in the super-insulating state with the superconducting phase localized within the separated grains [53]. The so-called quasi-reentrant behaviour develops by an interplay between the Josephson and quasiparticle tunneling across the low – and high-resistance junctions in weakly coupled superconducting systems [52]. Thus, the topology driven superconductor – insulator phase transition depends on the intergranular resistance in the way similar to the discussed here.

To summarize, we found the continuity percolation threshold in thin CoPd films at the onset of the long-range ferromagnetic order. The conductance percolation threshold was determined at the thickness at which the resistance diverged toward infinity by following the classical conductance scaling and the resistivity temperature coefficient reversed its polarity from a metal-like positive to an insulator-like negative at room temperature. The critical conductance thickness is about four times smaller than the continuity one. To explain this huge difference, we suggest the existence and phenomenological importance of the transition range between the topologically continuous and discontinuous phases in which narrow low resistance gaps intersect the fractal networks of metallic clusters. Resistance of these discontinuities immediately below the continuity threshold can be two orders of magnitude smaller than the quantum resistance value. The gaps expand with decreasing coverage and the conductance percolation threshold is interpreted as the point at which the resistance of intergranular junctions in discontinuous films exceeds the quantum resistance mark. Discontinuous films in the transition range mimic the continuous metallic behavior by the magnitude of resistivity, the positive room temperature resistivity temperature coefficient, and by following the classical metallic-like percolation conductivity scaling.

References

1. D. Stauffer and A. Aharony, Introduction to Percolation Theory (Taylor and Francis, London, 1994).

2. D. Stauffer, Scaling theory of percolation clusters, Physics Reports **54** 1 (1979).

3. B. Abeles, H. L. Pinch, and J. I. Gittleman, Percolation Conductivity in W-Al2O3 Granular Metal Films, Phys. Rev. Lett. 35, 247 (1975).

4. B. Abeles, P. Sheng, M. D. Coutts, and Y. Arie, Structural and electrical properties of granular metal films, Advances in Physics, 24, 407 (1975).

5. I. Balberg, Tunneling and nonuniversal conductivity in composite materials, Phys. Rev. Lett. 59 1305 (1987).

6. D. Toker, D. Azulay, N. Shimoni, I. Balberg, and O. Millo, Tunneling and percolation in metal-insulator composite materials, Phys. Rev. B **68**, 041403(R) (2003).

7. Z. Rubin, S.A. Sunshine, M.B. Heaney, I. Bloom, and I. Balberg, Critical behavior of the electrical transport properties in a tunneling-percolation system, Phys. Rev. B 59, 12196 (1999).

8. I. Balberg, Tunnelling and percolation in lattices and the continuum, J. Phys. D: Appl. Phys. 42 064003 (2009).

9. S. Fostner, R. Brown, J. Carr, and S. A. Brown, Continuum percolation with tunneling, Phys. Rev. B 89, 075402 (2014).

10. C. Grimaldi, Theory of percolation and tunneling regimes in nanogranular metal films, Phys. Rev. B 89, 214201 (2014).

11. P. Sheng, B. Abeles, and Y. Arie, Hopping Conductivity in Granular Metals, Phys. Rev. Lett. 31, 44 (1973).

12. A. Gerber, I. Kishon, D. Bartov, and M. Karpovski, Resistivity minimum in granular composites and thin metallic films, Phys. Rev. B 94, 094202 (2016) and references therein.

13. F. O. Schumann, M. E. Buckley, and J. A. C. Bland, Paramagnetic-ferromagnetic phase transition during growth of ultrathin Co/Cu(001) films, Phys. Rev. B 50, 16424 (1994).

14. E. Gu, S. Hope, M. Tselepi, and J. A. C. Bland, Two-dimensional paramagnetic-ferromagnetic phase transition and magnetic anisotropy in Co(110) epitaxial nanoparticle arrays, Phys. Rev. B 60, 4092 (1999).

15. D. Küpper, S. Easton, and J. A. C. Bland, Paramagnetic-ferromagnetic phase transition and magnetic properties of ultrathin CoFe/Cu(110) films, J. Appl. Phys. 102, 083902 (2007).

16. R. Belanger and D. Venus, Two-dimensional percolation transition at finite temperature: Phase boundary for in-plane magnetism in films with two atomic layers of Fe on W(110), Phys. Rev. B 95, 085424 (2017).

17. Y. B. Xu, E. T. M. Kernohan, D. J. Freeland, A. Ercole, M. Tselepi, and J. A. C. Bland, Evolution of the ferromagnetic phase of ultrathin Fe films grown on GaAs(100)-4 ×6, Phys. Rev. B 58, 890 (1998).

18. F. Bensch, G. Garreau, R. Moosbühler, G. Bayreuther, and E. Beaurepaire, Onset of ferromagnetism in Fe epitaxially grown on GaAs(001) (4×2) and (2×6), J. Appl. Phys. 89, 1 (2001).

19. Y. Shiratsuchi, T. Murakami, Y. Endo, and M. Yamamoto, Evolution of Magnetic State of Ultrathin Co Films with Volmer–Weber Growth, Jpn. J. Appl. Phys., Vol. 44, 8456 (2005).

20. S. Kirkpatrick, Percolation thresholds in Ising magnets and conducting mixtures, Phys. Rev. B 15, 1533 (1977).

21. C. Jayaprakash, E.K. Riedel, and M. Wortis, Critical and thermodynamic properties of the randomly dilute Ising model, Phys. Rev. B 18, 2244 (1978).

22. S. Bedanta and W. Kleemann, Supermagnetism, J. Phys. D: Appl. Phys. 42, 013001 (2009).

23. M.R. Scheinfein, K.E. Schmidt, K.R. Heim, and G.G. Hembree, Magnetic Order in Two-Dimensional Arrays of Nanometer-Sized Superparamagnets, Phys. Rev. Lett. 76, 1541 (1996).

24. V.N. Kondratyev and H.O. Lutz, Shell Effect in Exchange Coupling of Transition Metal Dots and Their Arrays, Phys. Rev. Lett. 81, 4508 (1998).

25. M. Varón, M. Beleggia, T. Kasama, R. J. Harrison, R. E. Dunin-Borkowski, V. F. Puntes, and C. Frandsen, Dipolar Magnetism in Ordered and Disordered Low-Dimensional Nanoparticle Assemblies. *Sci Rep* **3**, 1234 (2013).

26. R. M. Bozorth, P. A. Wolff, D. D. Davis, V. B. Compton, and J. H. Wernick, Ferromagnetism in Dilute Solutions of Cobalt in Palladium, Phys. Rev. 122, 1157 (1961).

27. H. Takahashi, S. Tsunashima, S. Iwata, and S. Uchiyama, Measurement of Magnetostriction Constants in (111)-Oriented Polycrystalline PdCo Alloy and Multilayered Films, Jpn. J. Appl. Phys. Part 2 32, L1328 (1993).

28. A. Segal, M.Karpovski, and A.Gerber, Sixteen-state magnetic memory based on the extraordinary Hall effect, J. Magn. Magn. Mat. 324 1557 (2012).

29. G. Kopnov and A. Gerber, Non-reciprocal magnetoresistance, directional inhomogeneity and mixed symmetry Hall devices, Appl. Phys. Lett. 119, 102405 (2021)

30. A. Gerber, G. Kopnov, and M. Karpovski, Hall effect spintronics for gas detection, Appl. Phys. Lett. **111**, 143505 (2017)

31. G. Kopnov and A. Gerber, Tuning the coercivity of ferromagnetic films with perpendicular anisotropy by thickness, width, and profile, Appl. Phys. Lett. 120, 182401 (2022).

32. A. Gerber, A. Milner, M. Karpovsky, B. Lemke, H.-U. Habermeier, J. Tuaillon-Combes, M. Negrier, O. Boisron, P. Melinon, and A. Perez, Extraordinary Hall Effect in Magnetic Films, J. Magn. Magn. Mat. 242, 90 (2002).

33. M. Volmer and A. Weber, Keimbildung in übersättigten Gebilden, *Zeitschrift für Physikalische Chemie* **119U,** 277 (1926).

34. M. Ohring, The materials science of thin films, Academic Press, Boston, MA (1992),

35. T. Harumoto, J. Shi and Y. Nakamura, Controllable magnetic anisotropy and magnetostriction constant in palladium cobalt alloy films: Effects of composition, thickness, and stress, J. Appl. Phys. 126, 083906 (2019).

36. T.R. McGuire and R.I. Potter, Anisotropic magnetoresistance in ferromagnetic 3d alloys, *IEEE Trans. Magn.* **11** 1018 (1975).

37. J. I. Gittleman, Y. Goldstein, and S. Bozowski, Magnetic Properties of Granular Nickel Films, Phys. Rev. B 5, 3609 (1972).

38. H. Fujimori, S. Mitani, and S. Ohnuma, Tunnel-type GMR in metal-nonmetal granular alloy thin films, Mater. Sci. Eng. B **31**, 219 (1995).

39. A. Milner, A. Gerber, B. Groisman, M. Karpovsky and A. Gladkikh, Spin-Dependent Electronic Transport in Granular Ferromagnets, Phys. Rev. Lett. 76, 475 (1996).

40. A. Gerber, A Milner, B. Groisman, M. Karpovsky, A. Gladkikh and A. Sulpice, Magnetoresistance of granular ferromagnets, Phys. Rev. B 55, 6446 (1997).

41. A. Gerber, B. Groisman, A. Milner and M. Karpovsky, Spin-dependent scattering in weakly coupled nickel films, Europhys. Lett. 49, 383 (2000).

42. B. Raquet, M. Viret, E. Sondergard, O. Cespedes, and R. Mamy, Electron-magnon scattering and magnetic resistivity in 3*d* ferromagnets, Phys. Rev. B **66**, 024433 (2002).

43. P. A. Lee and T. Ramakrishnan, Disordered electronic systems, Rev. Mod. Phys. **57**, 287 (1985).

44. B. L. Altshuler and A. Aronov, Electron–Electron Interaction In Disordered Conductors, in *Electron-Electron Interactions*, Disordered Systems Series, edited by A. L Efros and M. Pollak (Elsevier North Holland, Amsterdam, The Netherlands, 1985), Vol. 10, p. 690.

45. K. B. Efetov and A. Tschersich, Transition from insulating to non-insulating temperature dependence of the conductivity in granular metals, Europhys. Lett. **59**, 114 (2002).

46. K. B. Efetov and A. Tschersich, Coulomb effects in granular materials at not very low temperatures, Phys. Rev. B **67**, 174205 (2003).

47. J.H. Mooij, Electrical conduction in concentrated disordered transition metal alloys, Phys. Status Solidi (a) 17, 521 (1973)

48. J. W. Essam, C. M. Place and E. H. Sondheimer, Selfconsistent calculation of the conductivity in a disordered branching network, J. Phys. C, 7 (1974) L258.

49. M.S.M. Peterson and M. Deutsch, Conductivity and scaling properties of chemically grown granular silver films, J. Appl. Phys. 106, 063722 (2009).

50. B. G. Orr, H. M. Jaeger, A. M. Goldman, and C. G. Kuper, Global phase coherence in two-dimensional granular superconductors, Phys. Rev. Lett. **56**, 378 (1986).

51. S. Chakravarty, G.L. Ingold, S. Kivelson, and A. Luther, Onset of Global Phase Coherence in Josephson-Junction Arrays: A Dissipative Phase Transition, Phys. Rev. Lett. **56**, 2303 (1986).

52. S. Kobayashi and F. Komori, Phase coherence and critical resistance in a network of small Josephson junctions, J. Phys. Soc. Jpn 57, 1884 (1988).

53. A. Gerber, A. Milner, G. Deutscher, M. Karpovsky and A. Gladkikh, Insulator - Superconductor Transition in 3-D Granular Al - Ge Films, Phys. Rev. Lett. 78, 4277 (1997).

Figure captions

Fig.1. Transmission electron microscope images of 1.5 nm (a) and 7 nm (b) thick CoPd films. CoPd is dark in the figure. (c) High-resolution TEM image. Individual crystallites have fcc structure with pronounced (111) out-of-plane growth texture and random in-plane lattice orientations.

Fig.2. Normalized EHE resistance of CoPd films with different thicknesses at room temperature. Solid lines in films with thickness t ≤ 7 nm are fitting to the Langevin function (Eq.2). Inset: zoom of the 12.5 nm thick sample.

Fig.3. Planar resistance $R_\square$ (left vertical axis) and the effective magnetic moment $\mu$ (right vertical axis) as a function of thickness. $\mu$ was calculated by fitting the EHE data shown in Fig. 2 by the Langevin function (Eq.2). Red solid line is the fit of resistance by (Eq.4) with the critical thickness $t_{cR} = 2 \pm 0.2\ nm$ and $\gamma = 1.2 \pm 0.15$. Black solid line is fitting of magnetic moment $\mu$ by (Eq.2) with the critical thickness $t_{cM} = 7.5 \pm 0.2\ nm$ and power index $\alpha$ = -2.2. Dashed lines indicate the critical conductance $t_{cR}$ and ferromagnetic $t_{cFM}$ thicknesses respectively. Room temperature. The log-log presentation of $R_\square$ as a function of $t - t_{cR}$ is shown in the inset.

Fig. 4. Magnetoresistance of 4, 5.5, 6, 7, 8, and 10 nm thick films measured with field applied parallel to current. The magnetoresistance is normalized by the zero-field resistance. Room temperature.

Fig. 5. Planar resistance of several films in the thickness range $t_{cR} \leq t < t_{cFM}$ and a thicker 20 nm one as a function of the logarithm of temperature. The resistance is normalized by the room temperature value. Inset: $T_{min}$ as a function of thickness.

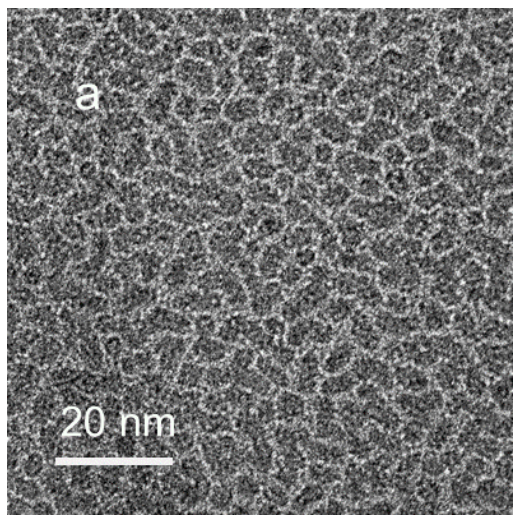
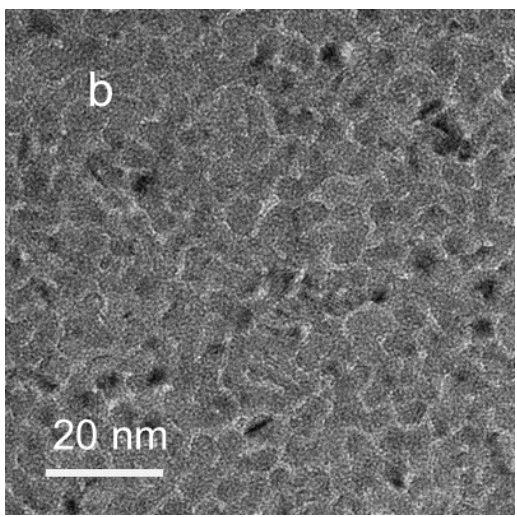
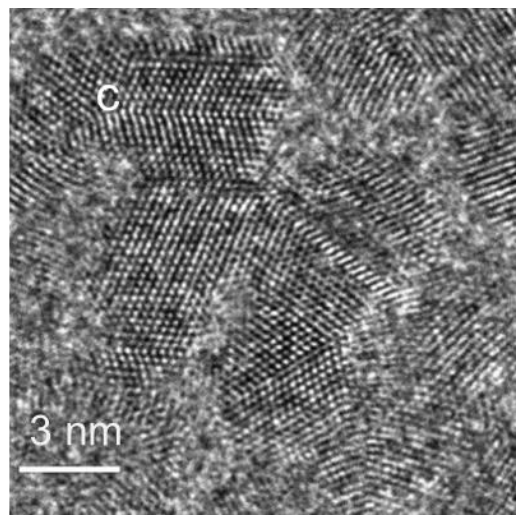

Fig. 1

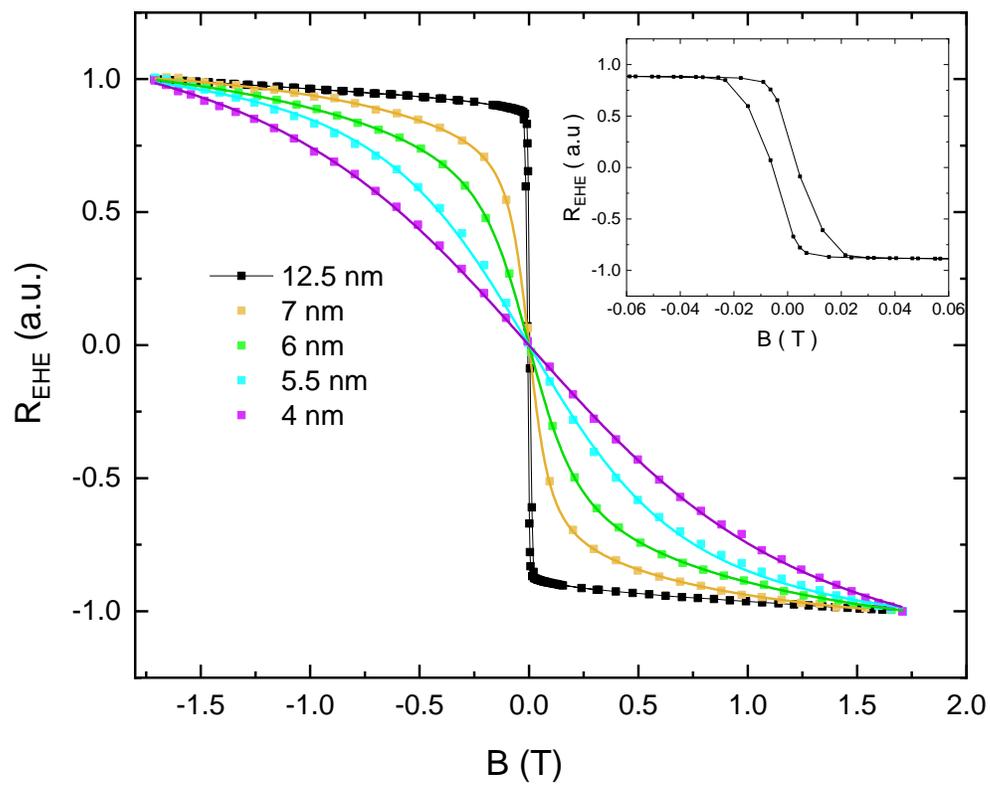

Fig. 2

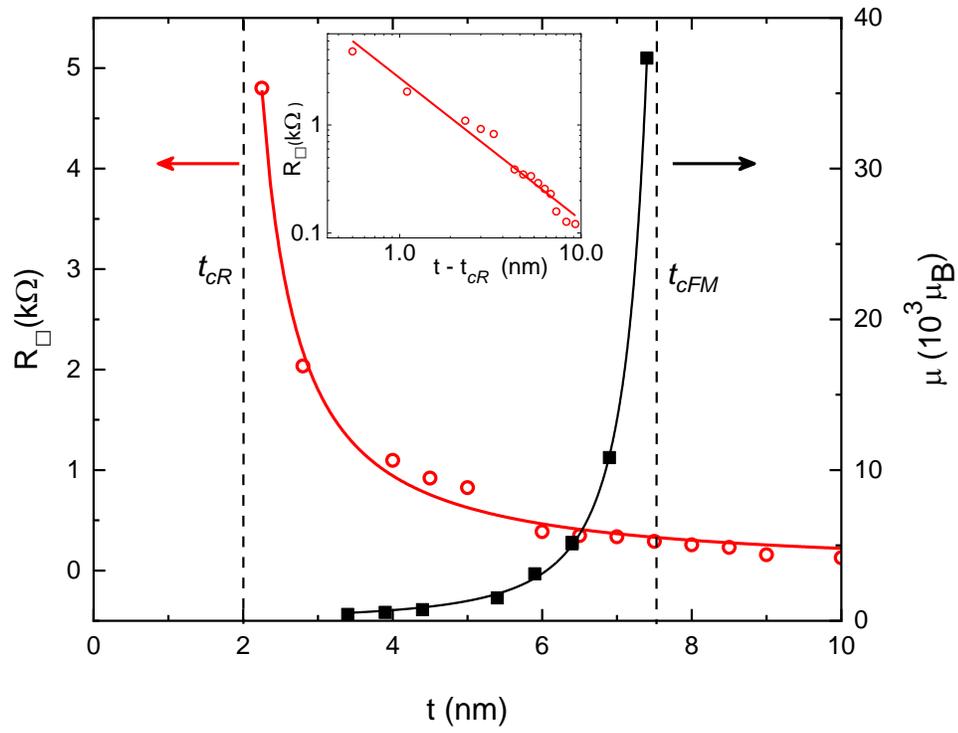

Fig. 3

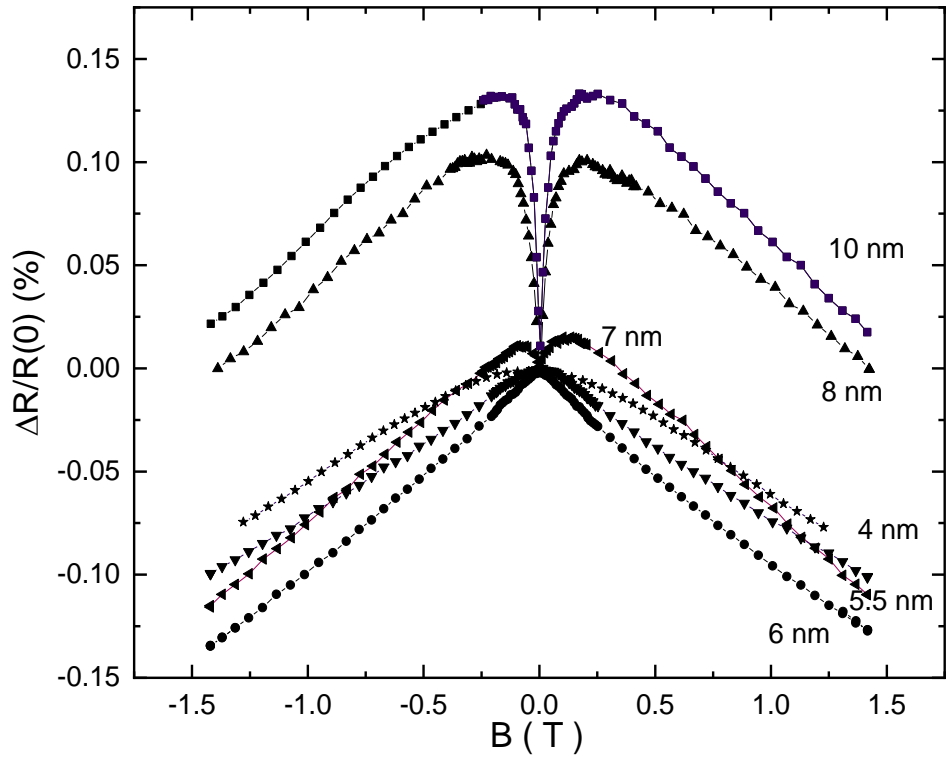

Fig. 4

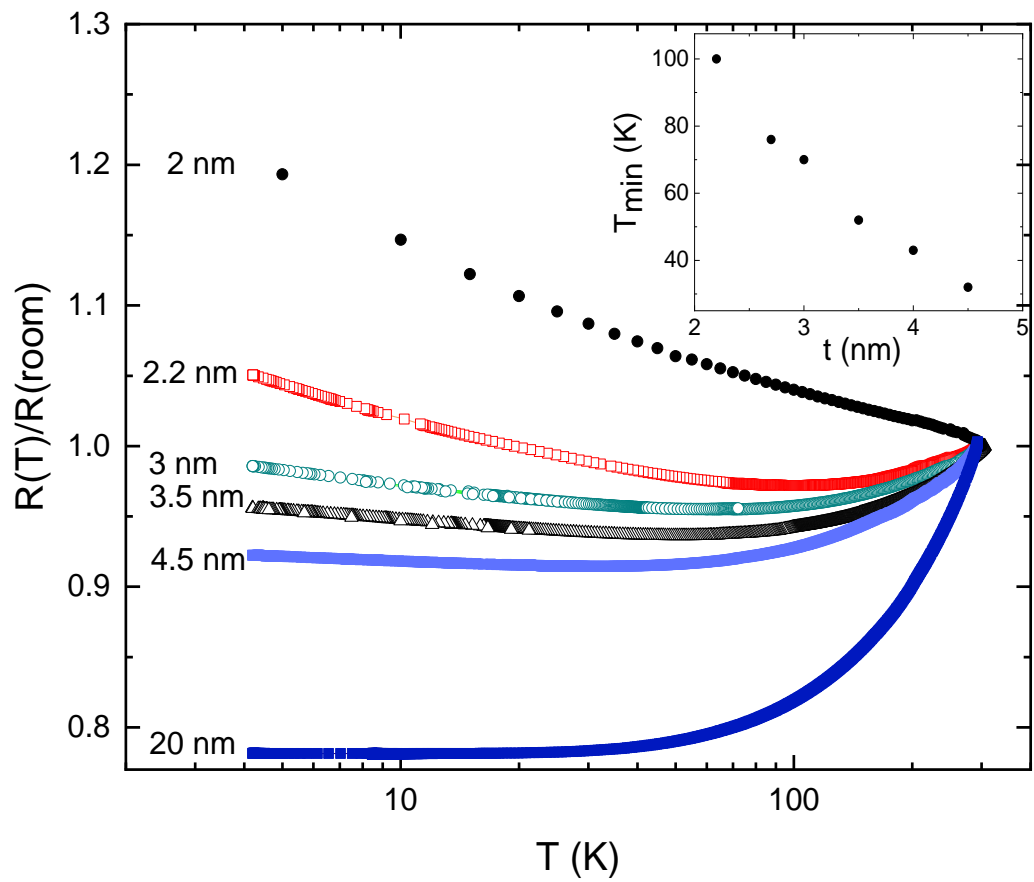

Fig. 5

20